%% file: 0-main.tex
\documentclass[manuscript]{aastex701}

\usepackage{epsfig}
\usepackage[version=4]{mhchem}

\shorttitle{Mars as an Exoplanet}
\shortauthors{Brain et al.}
\submitjournal{ApJ}
\graphicspath{{./}}

\begin{document}

\title{Atmospheric Escape Rates from Mars - If it Orbited an Old M-Dwarf Star}

\author[0000-0001-8932-368X]{David A. Brain}
\affiliation{Laboratory for Atmospheric and Space Physics, University of Colorado Boulder, USA}
\affiliation{Department of Astrophysical and Planetary Sciences, University of Colorado Boulder, USA}
\email{david.brain@lasp.colorado.edu}

\author[0000-0003-3721-0215]{Ofer Cohen}
\affiliation{Lowell Center for Space Science and Technology, University of Massachusetts Lowell, 600 Suffolk Street, Lowell, MA 01854, USA}
\email{ofer\_cohen@uml.edu}

\author[0000-0003-0912-8353]{Thomas E. Cravens}
\affiliation{Department of Physics and Astronomy, University of Kansas, Lawrence, KS, USA}
\email{cravens@ku.edu}

\author[0000-0002-1002-3674]{Kevin France}
\affiliation{Laboratory for Atmospheric and Space Physics, University of Colorado Boulder, USA}
\affiliation{Department of Astrophysical and Planetary Sciences, University of Colorado Boulder, USA}
\email{kevin.france@lasp.colorado.edu}

\author[0000-0001-9843-9094]{Alex Glocer}
\affiliation{NASA Goddard Space Flight Center, Greenbelt, MD USA}
\email{alex.glocer-1@nasa.gov}

\author[0000-0001-9504-0520]{Parker Hinton}
\affiliation{Laboratory for Atmospheric and Space Physics, University of Colorado Boulder, USA}
\email{parker.hinton@lasp.colorado.edu}

\author[0000-0002-5548-3519]{Francois Leblanc}
\affiliation{LATMOS/CNRS, Sorbonne University, Paris, France}
\email{francois.leblanc@latmos.ipsl.fr}

\author[0000-0003-2584-7091]{Yingjuan Ma}
\affiliation{Department of Earth, Planetary, and Space Sciences, University of California, Los Angeles, Los Angeles, CA, USA}
\email{yingjuan@igpp.ucla.edu}

\author[0000-0002-0998-0434]{Akifumi Nakayama}
\affiliation{Department of Physics, College of Science, Rikkyo University, Tokyo, Japan}
\email{anakayama@rikkyo.ac.jp}

\author[0000-0001-9135-2076]{Shotaro Sakai}
\affiliation{Faculty of Environment and Information Studies, Keio University, Fujisawa, Kanagawa, Japan}
\email{shotaro@sfc.keio.ac.jp}

\author[0000-0001-9038-6856]{Ryoya Sakata}
\affiliation{Research Center for Advanced Science and Technology, University of Tokyo, Tokyo, Japan}
\email{rsakata@g.ecc.u-tokyo.ac.jp}

\author[0000-0001-5557-9062]{Kanako Seki}
\affiliation{Research Center for Advanced Science and Technology, University of Tokyo, Tokyo, Japan}
\email{k.seki@eps.s.u-tokyo.ac.jp}


\author[0000-0001-5052-3473]{Juli\'{a}n D. Alvarado-G\'{o}mez}
\affiliation{Leibniz Institute for Astrophysics Potsdam, An der Sternwarte 16, 14482 Potsdam, Germany}
\email{julian.alvarado-gomez@aip.de}

\author[0000-0002-3321-4924]{Zachory Berta-Thompson}
\affiliation{Department of Astrophysical and Planetary Sciences, University of Colorado Boulder, USA}
\email{zach.bertathompson@Colorado.EDU}

\author[0000-0002-8548-4088]{Eryn M. Cangi}
\affiliation{Laboratory for Atmospheric and Space Physics, University of Colorado Boulder, USA}
\email{eryn.cangi@colorado.edu}

\author[0000-0002-1939-4797]{Michael Chaffin}
\affiliation{Laboratory for Atmospheric and Space Physics, University of Colorado Boulder, USA}
\email{michael.chaffin@LASP.colorado.edu}

\author[0000-0003-1931-124X]{Jean-Yves Chaufray}
\affiliation{LATMOS/CNRS, Sorbonne University, Paris, France}
\email{Jean-Yves.Chaufray@latmos.ipsl.fr}

\author[0000-0003-3290-818X]{Renata Frelikh}
\affiliation{Laboratory for Atmospheric and Space Physics, University of Colorado Boulder, USA}
\email{renata.frelikh@lasp.colorado.edu}

\author[0000-0002-7056-3517]{Yoshifumi Futaana}
\affiliation{Swedish Institute of Space Physics, Bengt Hultqvists väg 1, 981 92, Kiruna, Sweden}
\email{futaana@irf.se}

\author[0000-0001-6398-8755]{Katherine Garcia-Sage}
\affiliation{NASA Goddard Space Flight Center, Greenbelt, MD USA}
\email{katherine.garcia-sage@nasa.gov}

\author[0009-0001-8510-1729]{Lukas Hanson}
\affiliation{Lowell Center for Space Science and Technology, University of Massachusetts Lowell, 600 Suffolk Street, Lowell, MA 01854, USA}
\email{Lukas\_Hanson@student.uml.edu}

\author[0000-0001-5494-5374]{Mats Holmström}
\affiliation{Swedish Institute of Space Physics, Bengt Hultqvists väg 1, 981 92, Kiruna, Sweden}
\email{matsh@irf.se}

\author[0000-0002-0758-9976]{Bruce Jakosky}
\affiliation{Laboratory for Atmospheric and Space Physics, University of Colorado Boulder, USA}
\affiliation{Department of Earth and Space Sciences, University of Washington}
\email{bruce.jakosky@lasp.colorado.edu}

\author[0000-0002-4246-2954]{Riku Jarvinen}
\affiliation{Finnish Meteorological Institute, Helsinki, Finland}
\email{riku.jarvinen@fmi.fi}


\author[0000-0002-5893-2471]{Ravi Kopparapu}
\affiliation{NASA Goddard Space Flight Center, Greenbelt, MD USA}
\email{ravikumar.kopparapu@nasa.gov}

\author[0000-0001-6699-494X]{Daniel R. Marsh}
\affiliation{School of Physics and Astronomy, University of Leeds, Leeds, UK}
\email{D.Marsh@leeds.ac.uk}

\author[0000-0001-9751-6481]{Aimee Merkel}
\affiliation{Laboratory for Atmospheric and Space Physics, University of Colorado Boulder, USA}
\email{aimee.merkel@lasp.colorado.edu}

\author[]{Thomas Earle Moore}
\affiliation{3rd Rock Research, Scarborough, ME USA}
\email{thomem@iiirdrock.net}

\author[0000-0002-0412-0849]{Yuta Notsu}
\affiliation{Laboratory for Atmospheric and Space Physics, University of Colorado Boulder, USA}
\affiliation{National Solar Observatory}
\affiliation{Department of Astrophysical and Planetary Sciences, University of Colorado Boulder}
\email{yuta.notsu@lasp.colorado.edu}

\author[0000-0001-5643-8421]{Rachel A. Osten}
\affiliation{Space Telescope Science Institute}
\affiliation{Johns Hopkins University Center for Astrophysical Sciences, Baltimore MD 21218}
\email{osten@stsci.edu}

\author[]{William K. Peterson}
\affiliation{Laboratory for Atmospheric and Space Physics, University of Colorado Boulder, USA}
\email{William.Peterson@lasp.colorado.edu}

\author[0000-0002-4438-3504]{Laura Peticolas}
\affiliation{EdEon, Sonoma State University, 1801 E Cotati Ave, Rohnert Park 94928, United States}
\email{laurap@universe.sonoma.edu}

\author[0000-0003-0458-4050]{Robin Ramstad}
\affiliation{LASP, University of Colorado Boulder}
\email{robin.ramstad@lasp.colorado.edu}

\author[0000-0002-7352-7941]{Kevin B. Stevenson}
\affiliation{JHU Applied Physics Laboratory, 11100 Johns Hopkins Rd, Laurel, MD 20723, USA}
\email{Kevin.Stevenson@jhuapl.edu}

\author[0000-0001-9839-1828]{Robert Strangeway}
\affiliation{Department of Earth, Planetary, and Space Sciences, University of California, Los Angeles, Los Angeles, CA, USA}
\email{strange@igpp.ucla.edu}

\author[0000-0001-6699-7585]{Wenyi Sun}
\affiliation{Department of Climate and Space Sciences and Engineering, University of Michigan, Ann Arbor, MI, USA}
\email{wenyisun@umich.edu}

\author[0000-0001-5685-9736]{Naoki Terada}
\affiliation{Department of Geophysics, Graduate School of Science, Tohoku University, Sendai, Japan}
\email{teradan@pat.gp.tohoku.ac.jp}

\author[0000-0001-5371-2675]{Aline A. Vidotto}
\affiliation{Leiden Observatory, Leiden University}
\email{vidotto@strw.leidenuniv.nl}

\begin{abstract}

Atmospheric escape is an important process that influences the evolution of planetary atmospheres. A variety of physical mechanisms can contribute to escape from an atmosphere, including thermal escape, ion escape, photochemical escape, and sputtering. Here we estimate escape rates via each of these processes for a hypothetical Mars-like exoplanet orbiting Barnard's star (an old, inactive M dwarf star). We place the planet at an orbital distance that receives the same total stellar flux as it does in our solar system. We use the measured stellar extreme ultraviolet (EUV) spectrum and assumptions on the star's magnetic field to determine both the high-energy radiation and the stellar wind environment around the planet. This information is used to model the response of the planet's thermosphere, exosphere and magnetosphere using a variety of models that have been validated against solar system observations. We find overall escape rates that are dominated by thermal processes and elevated by 2-5 orders of magnitude relative to present-day Mars, suggesting that a Mars-like planet orbiting Barnard's star would not retain a significant atmosphere for more than 10's of millions of years. Recently reported planets around Barnard's star should also not have retained significant atmospheres. By extension, Mars-like planets orbiting any M dwarf near the 'Habitable Zone' should not retain atmospheres for extended periods of time.

\end{abstract}



\input{1-intro.tex}

\input{2-approach.tex}

\input{3-inputs.tex}

\input{4-thermosphere.tex}

\input{5-hydrodynamic.tex}

\input{6-ion.tex}

\input{7-photochemical.tex}

\input{8-sputtering.tex}

\input{9-synthesis.tex}

\input{10-discussion.tex}

\begin{acknowledgments}
This work was funded by NASA ICAR grant 80NSSC23K1358 (Retention of Habitable Atmospheres)and NASA DRIVE grant 80NSSC20K0594. The authors acknowledge the MIT SuperCloud and Lincoln Laboratory Supercomputing Center for providing HPC resources that have contributed to the research results reported within this paper. AAV acknowledges funding from the European Research Council (ERC) under the European Union's Horizon 2020 research and innovation programme (grant agreement No 817540, ASTROFLOW) and funding from the Dutch Research Council (NWO), with project number VI.C.232.041 of the Talent Programme Vici. KS acknowledges NASA ICAR Grant No. 80NSSC23K1399 (Strange New Worlds). R.J. received funding from the European Research Council (Grant
agreement No. 101124960).
\end{acknowledgments}


\bibliography{bib}{}
\bibliographystyle{aasjournalv7}



\end{document}

%% file: 1-intro.tex
\section{Introduction} \label{sec:intro}

As planetary atmospheres form and evolve, their mass at any given moment is determined by the time history of the source and loss processes that have acted on them. 
An important (and permanent) loss process is atmospheric escape to space. Atmospheric escape has been measured or indirectly inferred from a large number of solar system objects \citep{anderson1971, thomas1972, broadfoot1986, brace1987, hartle2006, gladstone2016}. Atmospheric isotope signatures further suggest that loss processes have played an important role in the evolution of solar system atmospheres \citep{owen1977, donahue1982, jakosky1997}. Observations from the Mars Atmosphere and Volatile EvolutioN (MAVEN) mission indicate that Mars has lost at least half a bar of atmosphere over its history \citep{jakosky2018}.

Atmospheric escape is known to be an active process at exoplanets, with multiple observations of strong escape from gas giant planets \citep[e.g.][]{vidalmadjar2003, lecavelier2010, kulow2014, ehrenreich2015, spake2018}. Sustained atmospheric escape is inferred to be the main reason for the `radius gap', or deficit of observed exoplanets orbiting close to their star with sizes between Super-Earth and Sub-Neptune \citep{owen2017, fulton2017}. There are now multiple observations of rocky exoplanets lacking a substantial atmosphere \citep[e.g.][]{kreidberg2019, crossfield2022}, implying that escape has removed their atmospheres. The James Webb Space Telescope (JWST) has enabled high spectral resolution transit spectroscopy of rocky exoplanets, with further evidence for rocky planets lacking significant atmospheres \citep[e.g.][]{greene2023,wachiraphan2025}. The JWST-HST (Hubble Space Telescope) Rocky Worlds Director's Discretionary Time (DDT) program is dedicated to a search for atmospheres on rocky exoplanets \citep{redfield2024}. Understanding of atmospheric escape can be used to interpret these observations, and place them in context with each other and with solar system planets. They can also be used to test the proposed `cosmic shoreline' framework for evaluating whether a given planet will retain an atmosphere \citep{zahnle2017}. 

The term `atmospheric escape' refers to a suite of processes, rather than a single mechanism that removes particles from a planet's atmosphere. Different investigators identify different numbers of escape processes, though this is an issue mostly of classification. There is general agreement on the physics that governs escape \citep[e.g.][]{hunten1982, brain2016, gronoff2020}. Here we identify five main processes, following \citet{brain2016}: thermal Jeans escape, hydrodynamic escape, ion escape, photochemical escape, and sputtering. Thermal and hydrodynamic escape both result from the thermal distribution of particle energies in a planet's upper atmosphere. Some fraction of this distribution can exceed the escape energy from the planet, resulting in escape. If a significant fraction of the distribution exceeds the escape energy then the resulting escape can be characterized as a fluid outflow, and is referred to as hydrodynamic escape. Ion, photochemical, and sputtering escape are all non-thermal processes. In these cases some external energy (electric fields, exothermic reactions, and particle collisions, respectively) is required to accelerate particles away from the atmosphere. Charge exchange (between stellar wind protons and exospheric neutrals) is sometimes considered as a separate process; in this work it is included as a chemical pathway that leads to ion escape. Different escape processes are effective at removing different atmospheric species, and the importance of individual processes can change as a planet and its host star change with time.

Many previous studies of atmospheric escape from exoplanets have included hydrodynamic escape as the dominant relevant escape process. Hydrodynamic escape is easiest to initiate for low-mass species such as hydrogen or helium, and previous studies have considered hydrodynamic escape from giant planets \citep[see a reviews in][]{owen2019, hazra2025} and from rocky planets with their primordial H/He atmospheres \citep[e.g.][]{lehmer2017, ito2021}. Energy-limited hydrodynamic escape fluxes are straightforward to evaluate using known parameters of the planet (mass, radius, incident stellar flux at EUV/X-ray wavelengths), though they include efficiency parameters that are difficult to constrain \citep[e.g.][]{erkaev2007}. 

Other exoplanet studies have considered the influence of non-thermal processes from exoplanets. These include estimates of photochemical escape \citep{hu2012, lee2021} and of ion escape \citep[e.g.][]{garciasage2017, airapetian2017, dong2017a, dong2017b, dong2018, dong2019, dong2020, egan2019}. 

Comparatively few studies have considered the net atmospheric escape from exoplanets via multiple escape processes, despite the likelihood that multiple escape processes operate on many exoplanets, as they do on the planets in our solar system. Examples include \citet{france2020} and \citet{avtaeva2022}. To our knowledge, no studies have considered all five escape processes from an exoplanet. 

Here we present a comprehensive modeling case study of a Mars-like exoplanet orbiting an M dwarf star. We choose a Mars-like planet because four of the five escape processes operate on Mars today (all but hydrodynamic escape) and because observations of atmospheric escape are currently more complete at Mars than they are at Venus or Earth \citep{jakosky2018}. We place Mars in its present state around an M dwarf star because M dwarfs make up $\sim$75\% of the stars in the Galaxy and because rocky exoplanet planet characterization is only possible around M dwarfs using current facilities (i.e., JWST). M dwarf stars are known to remain particularly magnetically active for a substantial fraction of their lives. This activity seems likely to drive enhanced atmospheric escape - an idea supported by the absence of atmospheric detections on planets orbiting M dwarfs to this point \citep{kreidberg2025}. Our goals with this study are (1) to demonstrate a methodology for comprehensively estimating atmospheric escape rates, (2) to determine how long present-day Mars might retain its atmosphere if it orbited an M dwarf star at a location where it receives the same total stellar flux, and (3) to highlight caveats associated with these and other published estimates of atmospheric escape that we believe are important for the community to keep in mind.

%% file: 2-approach.tex
\section{Approach} \label{sec:approach}

In this study we estimate escape rates from Mars if it orbited an M dwarf star. Our goal is to use models to determine whether atmospheric escape rates via each of the five main escape processes would increase or decrease (relative to present-day Mars) in this situation, and by how much. To prevent confusion, note that we use the terms `present-day Mars' and `exo-Mars' to refer to Mars in the present-day solar system and our modeled Mars orbiting Barnard's star, respectively.

We use an end-to-end modeling approach, starting with the stellar inputs for atmospheric escape, and modeling both the upper atmospheric regions relevant for atmospheric escape as well as the escape processes themselves. We first choose relevant planetary and stellar parameters (Section 3) that serve as inputs to models for the planet's upper atmosphere, magnetosphere, and escape processes. Next, we use these input parameters to model the exo-Mars thermosphere subject to photon fluxes from the star (Section 4). Next, we use the modeled thermosphere to evaluate hydrodynamic and thermal escape rates (Section 5). Three models for the magnetosphere and ion escape are applied, using the modeled thermosphere as the lower boundary input to the simulations (Section 6). We use the modeled thermosphere and ionosphere to estimate photochemical escape rates (Section 7), and the magnetosphere model to estimate sputtering rates (Section 8). Where possible we use multiple models or computation methods in order to provide an indication of uncertainty in our estimated escape rates.

Section 9 synthesizes our results and places them in context with present-day Mars. Section 10 presents important caveats associated with our exercise, which may suggest how past and future efforts to estimate escape should be conducted and interpreted.

%% file: 3-inputs.tex
\section{Input Conditions} \label{sec:inputs}

\subsection{Planetary Parameters} \label{subsec:planet_inputs}

We have chosen our exo-Mars modeling case to be identical to present-day Mars in terms of its basic planetary parameters, with a planetary mass of $6.4 \times 10^{23}$ kg, radius of 3400 km, and a bulk atmosphere comprised of $\mathrm{CO_2}$. We model present-day Mars (i.e. a thin, $\sim$7 mbar atmosphere) as opposed to a thicker atmosphere because this state provides the upper atmospheric boundary conditions for our models that we well measured. We place exo-Mars at an orbital distance from the star (0.087 AU) that experiences the same stellar flux that preset-day Mars experiences in our own solar system.

\subsection{Stellar EUV Spectrum} \label{subsec:euv_inputs}

We utilize the stellar input spectrum of Barnard's star presented by~\citet{france2020}, obtained as part of the Mega-MUSCLES observing program \citep{wilson2025}. Barnard's Star is an M3V star with approximately 0.16~M$_{\odot}$~\citep{ribas18}. It has an estimated age of 7~--~12 Gyr, based on a combination of slow rotation period ($P_{rot}$ = 130~--~145 days; \citealt{benedict1998, toledo2019}), low X-ray luminosity~\citep{stelzer13,guinan19}, and low magnetic activity levels~\citep{hunsch99,ribas18}. The advanced age and very low intrinsic UV and X-ray luminosity for an M dwarf star~\citep{france2020} make the present study a conservative lower limit to the mass loss rates expected for Martian planets orbiting mid-M type stars. While Barnard's Star has been observed to flare at UV and X-ray wavelengths \citep{france2020}, we only consider the quiescent photon continuum in this work. For reference, the integrated UV and X-ray flux from Barnard's star is $\sim~3\times$ that from the Sun. 

\begin{figure*}
\begin{center}
\hspace{-0.0in}\epsfig{figure=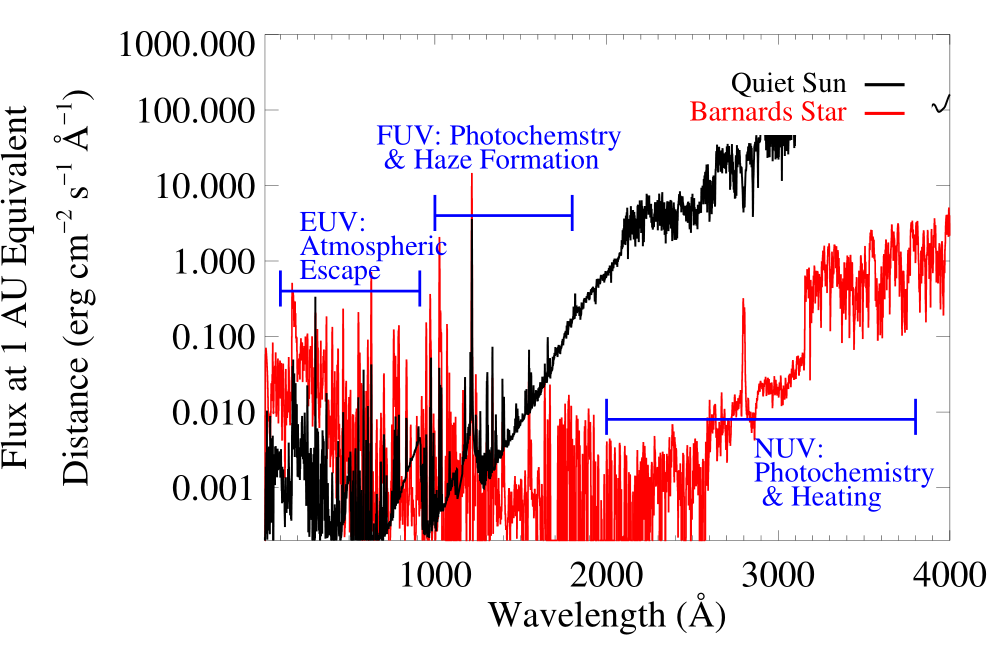,width=5in}
\vspace{+0.0in}
\caption{
\label{cosovly} 
The quiescent spectrum of Barnard's Star (GJ 699, shown in red), is shown in comparison with the spectrum of the quiet Sun (from \citealt{woods2009}, shown in black).  These spectra are shown at 2~\AA\ resolution and scaled to a common bolometric instellation distance (Figure adapted from \citealt{france2020}). The EUV flux is enhanced relative to the quiet Sun by a factor of 2-10, even for this inactive M dwarf, due to the enhanced EUV/bolometric fraction in low mass stars and the smaller orbital radius of the habitable zone around lower luminosity M dwarf stars
}
\label{fig:euv} 
\end{center}
\end{figure*}

The Barnard's Star spectral energy distribution used in this work is based on FUV (Far UltraViolet) emission line observations with the {\it Hubble Space Telescope} and X-ray spectra from the {\it Chandra X-ray Observatory}.  These spectra provide emission originating in the upper stellar atmosphere, from the chromosphere to the corona~(formation temperature~$\sim$~10$^{4}$~--~10$^{7}$ K).  The temperature-dependent fluxes are used to develop the emission measure of the stellar atmosphere (see, e.g., \citealt{kashyap1998,sanzforcada2013}), which can be fit by a smooth function to estimate the distribution of gas that contributes the majority of the EUV flux in stellar atmospheres ($\sim$~10$^{5.2}$~--~10$^{6.2}$ K).  The emission measure distribution is coupled with an emission contribution function to predict the currently unobserved EUV (100 - 912\AA) flux~\citep{duvvuri2021} that is used as an input for the present modeling study.  Figure 1 displays the EUV through optical spectrum of Barnard's star and the quiet Sun, normalized to a common instellation distance. This figure illustrates the increased EUV flux incident on planets orbiting even inactive M dwarfs compared to the Mars-Sun system.

\subsection{Stellar wind and interplanetary magnetic field} \label{subsec:swimf_inputs}

The stellar wind conditions of Barnard's Star are the same as those calculated by \cite{alvaradogomez2019}. These conditions were obtained using the AWSOM (Alfvén Wave SOlar Model)) Magnetohydrodynamic (MHD) model \citep{vanderholst2014}, which is part of the Space Weather Modeling Framework \citep[SWMF,][]{toth2012,gombosi2018}. The model calculates the steady-state MHD solution for the stellar corona and stellar wind, taking into account the effect of Alfv\'en waves on coronal heating and wind acceleration, as well as thermodynamic effects, such as electron heat conduction and radiative cooling. 

The model is driven by observations of the radial stellar magnetic field (magnetograms), which are imposed on the model's inner boundary. This boundary condition is used to obtain the initial, three-dimensional potential field of the star. The initial condition is then driven to a non-potential, MHD steady-state by imposing the plasma forcing on the magnetic field via the MHD equations, which include the plasma dynamics and thermodynamics. This steady-state MHD solution provides the values of the MHD parameters - density, velocity vector, magnetic field vector, and pressure anywhere in the simulation domain, which covers the space between the star's lower corona and (typically) a few tens of stellar radii.

The AWSOM model is driven by observations of the stellar photospheric field. However, such data for stars is extremely limited and is not available for Barnard's Star. \cite{alvaradogomez2019} used the star HD 179949, which does have such data, as a proxy star for Barnard's Star, arguing that stars with a similar Rossby number could serve as a good proxies due to a similarity in their magnetic activity (the Rossby number captures both the rotation period, the size of the star, and indirectly, the star's age). This approach has been validated recently by \cite{garraffo2022}, who confirmed that models derived from a proxy-star magnetogram data and an actual observed magnetogram data for Proxima Centauri were in reasonable agreement.  

\begin{figure}[htb]
\centering
\includegraphics[width=0.5\linewidth]{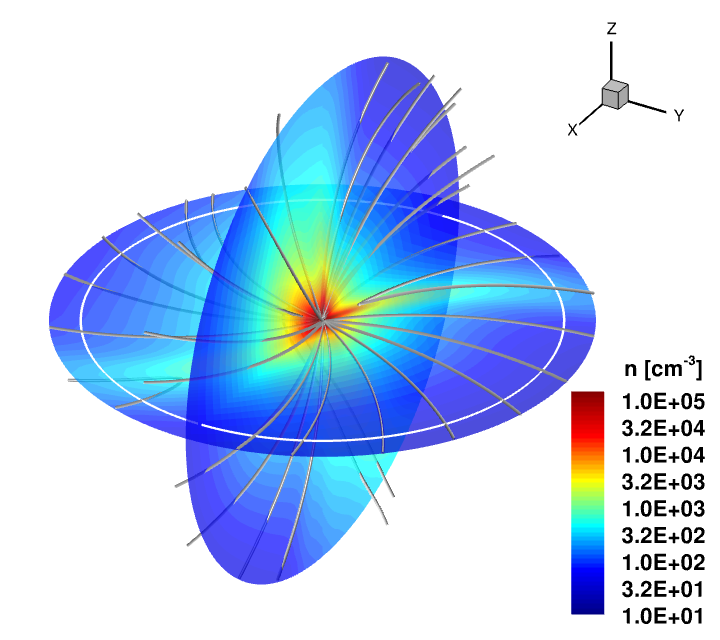}
\caption{The three-dimensional MHD stellar wind solution for Barnard's Star. Color contours represent number density values, while selected magnetic field lines are also shown. The solid white circle marks the orbit of the planet. \label{fig:BS_3d}}
\end{figure}

\begin{figure}[htb]
\centering
\includegraphics[width=0.6\linewidth]{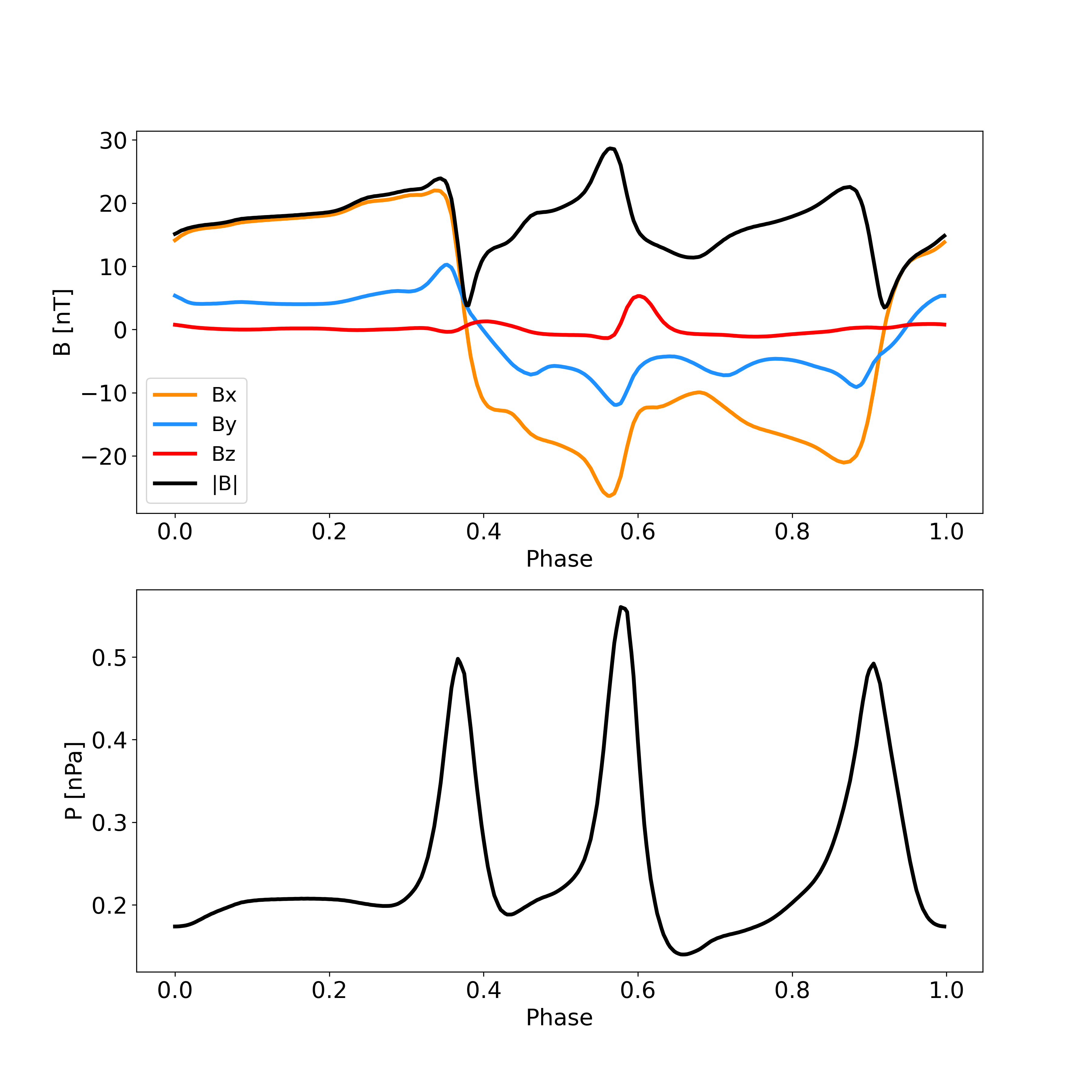}
\caption{Top: the value of the IMF components and the IMF magnitude along the circular orbit at 0.087 AU. Bottom: the stellar wind dynamic pressure along the circular orbit at 0.087 AU. The phase is determined with respect to the temporal longitude as defined by the observed magnetogram and the rotation of the star as observed from the Earth. This also defines the coordinate system of the domain.\label{fig:imf}}
\end{figure}

Figure~\ref{fig:BS_3d} shows the three-dimensional MHD solution for Barnard's Star. Figure~\ref{fig:imf} shows the value of the different components of the Interplanetary Magnetic Field (IMF), and its magnitude along the orbit of our hypothetical exoplanet at 0.087 AU. It also shows the dynamic pressure of the stellar wind along the orbit.

The stellar wind parameters at the location of the planet (used as input to all three magnetosphere models described in Section \ref{sec:ion}) are calculated by averaging the MHD results along the orbit of the planet. This yields a stellar wind density $\mathrm{N_{SW}}$ = 67.5 $\mathrm{cm^{-3}}$, velocity $\mathrm{\vec{U}_{SW}}$ = (-602.1, 0.0, 0.01) km/s, Interplanetary magnetic field $\mathrm{B_{IMF}}$ = (-15.4,-4.9, 0.04) nT, and temperature $\mathrm{T_{SW}}$ =  $\mathrm{4.8 \times 10^5}$ K. For reference, $\mathrm{N_{SW}}$ at present-day Mars is $\sim$ 1-5 $\mathrm{cm^{-3}}$, $\mathrm{U_{SW}} \sim (200-800, 0, 0)$ km/s, $\mathrm{B_{IMF}} \sim (1.2-3.1, 1.6-3.9,0)$ nT (corresponding to 2-5 nT with a spiral angle of $52^\circ$), and $\mathrm{T_{SW}} \sim 10^4 - 10^5$ K. Thus exo-Mars experiences higher stellar wind density, comparable stellar wind velocity, larger interplanetary magnetic field strength, and higher stellar wind temperatures.

%% file: 4-thermosphere.tex
\section{Modeled thermosphere} \label{sec:thermosphere}

We employ a one-dimensional thermospheric model to estimate the temperature and compositional structures of the exo-Mars thermosphere for given planetary parameters and stellar inputs. 
The model also provides the ionization frequencies which are used to evaluate ion escape as discussed in Section~\ref{sec:ion}. 
Our model is based on the model developed in \citet{Nakayama2022}  and slightly modified for application to CO$_2$ atmospheres and low-mass planets, which has been validated by comparison with MAVEN observations \citep[submitted to JGR]{Nakayama2025}. 
A brief summary of the model is provided below, while detailed model descriptions and applications are available in \cite{Nakayama2022} and \citet[submitted to JGR]{Nakayama2025}.

The model integrates the time-dependent continuity and energy equations from the bottom boundary (100 km in this case) up to the exobase. 
The model solves the steady-state momentum equation given the outflow velocity at the exobase to compute the temperature structure, as in \citet[]{Tian2008}.
Because the model does not solve the equation of motion, we adopted a density-weighted Jeans effusion velocity to represent the mean outflow velocity.
To stabilize the calculation of a weakly bounded atmosphere, we also adopt the mean outflow velocity to estimate the escape rate of each species.  
This is because, in the small Jeans escape regime, using the individual Jeans escape flux as outflow flux leads to a steep compositional gradient at the upper boundary.
This gradient arises from the discrepancy between outflow speed and the mean upward velocity or diffusional velocity. 
The physical processes included in our model are thermochemistry and photochemistry, molecular and atomic radiative cooling, molecular and eddy diffusion, and advection.
Endothermic reactions are included as the reverse of exothermic reactions to accommodate high-temperature conditions.
Note that the majority of reactions are exothermic in low-temperature conditions, as is the case with the planets in the solar system.
This model already includes the species composed of H, C, O, and N, following \citet{Nakayama2022}. 
In addition, we include species composed of He and Ar for a Mars-like atmosphere: He, Ar, ArH, He$^+$, Ar$^+$, and HeH$^+$.
Chemical reactions involving noble gases follow the chemical network given in \citet{Johnstone2018}.

Eddy diffusivity is an uncertain parameter for modeling the 1D thermosphere, because it is closely related to the CO$_2$ radiative damping of gravity waves \citep[]{Eckermann2011, Terada2017}. The estimated magnitude and altitude dependence of the eddy diffusivity have been controversial even for solar system planets \citep[e.g., ][]{vonZahn1980, Mahieux2021, Yoshida2022}. 
For simplicity, we adopt an altitude-independent value of $1.0 \times 10^7$~cm$^2$/s.
This value is consistent with the estimate for present-day Mars \citep[]{Yoshida2022} and present-day Venus \citep[]{Mahieux2021}.
The lower boundary of the model is set at 100~km, where present-day Martian conditions are deduced from the Viking mission \citep[]{McElroy1977, Fox1979, Shinagawa1989}.
In addition to the XUV spectrum of Barnard's star described in Section~\ref{sec:inputs}, we adopt an effective temperature of 3134~K for the infrared absorptions \citep[]{Dawson2004}. 
The atmospheric structure is calculated under angle-averaged conditions, assuming a solar zenith angle of 66$^\circ$, which is representative of a globally averaged profile for Earth \citep{Johnstone2018}. 

\begin{figure*}[t]
\begin{center}
\includegraphics[width=0.95\linewidth]{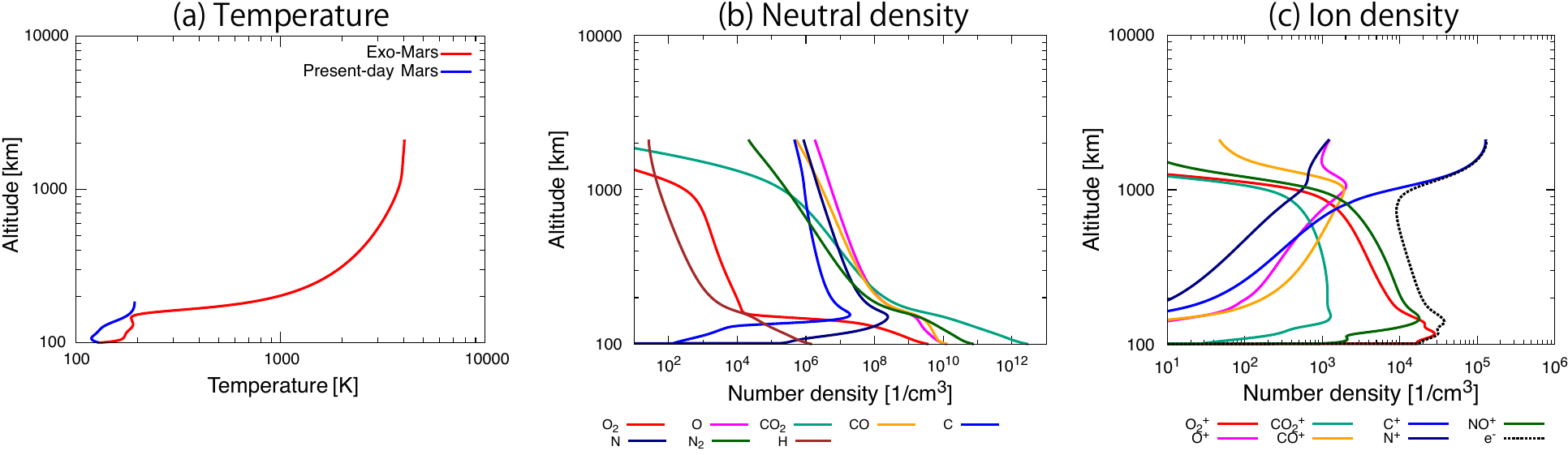}
\end{center}
\caption{(a)Temperature, (b)Neutral density, and (c)Ion density profiles of the modeled thermosphere for Exo-Mars. In the temperature profile, red and blue line indicate profiles of Exo-Mars and present-day Mars, respectively.}
\label{Fig:thermosphere}
\end{figure*}

Figure~\ref{Fig:thermosphere} shows modeled profiles of (a) temperature, (b) neutral density, and (c) ion density. In the temperature profile, we also show the profile of present-day Mars estimated from the blackbody spectrum of 5777~K for infrared absorption and XUV spectrum of FISM-M \citep{Thiemann2017}. The thermosphere is significantly expanded compared to the present-day Mars because of the intense XUV flux from Barnard's star relative to the Sun. The exobase altitude increases from 210 km \citep{Fu2020} to $>2000$ km. The shape of the temperature structure, except for the lower thermosphere, resembles that of \ce{CO2}-dominated atmospheres on solar system planets. The lower thermosphere is strongly heated by near-infrared absorption of \ce{CO2}, as lower stellar effective temperatures result in greater emission at longer wavelengths. The temperature at the exobase (at an altitude of approximately 2100~km) is $\sim$4000~K, which is significantly higher than that of 180-260~K \citep{Bougher2023} for present-day Mars and also exceeds the values in a previous study assuming 10 times the present-day solar XUV flux \citep{Tian2009}. This is due to the inhibition of \ce{CO2} radiative cooling and the low molecular weight of the upper thermosphere. \ce{CO2} is dissociated by photons with wavelengths shorter than $200$~nm.  In particular, photons with a wavelength between 100 and 120~nm significantly enhance the photodissociation of \ce{CO2}. This is because of the stronger emission from Barnard's star at these wavelengths compared to the solar spectrum shown in Fig.~\ref{fig:euv}, as well as the high transparency of these photons in the upper atmosphere enriched with atoms.  Because atomic radiative cooling becomes effective at temperatures larger than $3000$~K \citep{Nakayama2022}, it operates around the exobase. However, in the case of low-mass planets such as Mars, conduction and advective cooling are the dominant cooling processes. 

The combination of high temperature and low gravity promotes thermal escape. Global Jeans escape rates of major neutral species are shown in Table~\ref{Tab:ionizationfrequency}. Note that we ignored the effects of the truncated velocity distribution at the exobase due to the loss of energetic atoms, which reduces the actual escape flux \citep{Chaufray2021}.
For present-day Mars, the low dayside exobase altitude and the lower exospheric temperature ranges from 180 to 260~K indicate that the Jeans escape process is thus only effective for light species on present-day Mars, such as hydrogen.

Under strong FUV irradiation, efficient CO$_2$ dissociation leads to an upper thermosphere enriched with various species. 
While O remains the most abundant species similar to present-day Mars, N, C, and CO are also found in substantial abundances. 
In particular, C is typically found in strong UV environments because it is primarily produced through a two-step dissociation process from CO$_2$:
\begin{eqnarray}
    \mathrm{CO_2} &+& h\nu \rightarrow \mathrm{CO + O} \\
    \mathrm{CO} &+& h\nu \rightarrow \mathrm{C + O} .
\end{eqnarray}
The ion composition differs significantly from that of present-day Mars, as the ion profiles do not directly mirror the neutral profiles.
In the upper thermosphere, \ce{C+} is the most abundant ion species and plays a critical role in shaping the overall ion composition.
Table~\ref{Tab:ionizationfrequency} summarizes the ionization frequencies of major species and their resulting escape rates. Note that H is the only species that escapes in measurable quantities from present-day Mars, at rates \citep[$10^{26}-10^{27}~\mathrm{s^{-1}}$; ][]{chaffin2018} much higher than the rates for exo-Mars. A larger variety of species escape thermally from exo-Mars, and the total Jeans escape rate (summed over all species) is comparable to or greater than that for present-day Mars.
Since the ionization frequency of C is similar to that of O, C$^+$ is not primarily produced by the photochemistry. 
Instead, \ce{C+} is mainly generated via charge-exchange between \ce{O+} and C \citep{Fox2005}:
\begin{equation}
    \mathrm{C + O^+} \rightarrow \mathrm{C^+ + O} .
\end{equation}
The rate coefficient for the charge-exchange reaction is approximately 200 times greater than that of \ce{O+} recombination at 4000~K. 
As a result, C$^+$ becomes the dominant ion species instead of \ce{O+} under intense UV irradiation, where \ce{CO2} is efficiently dissociated.

In contrast to low-UV environments such as Mars and Venus, where \ce{O+} is the primary ionospheric species, \ce{O+} is rapidly depleted above 1000~km through the charge-exchange reaction. 
\ce{N+} is also produced via charge-exchange between \ce{O+} and N. 
However, the rate coefficient is only about 2\% of that of the rate coefficient for the reaction of \ce{O+} with C. Consequently, \ce{N+} remains a minor species and its abundance is closely related to C-N-O ion chemistry. \ce{H+} is also a very minor species below the exobase for exo-Mars, with number densities less than $10^{-1} cm^{-3}$.

Our results indicate the absence of a distinct electron density peak within the ionosphere below the exobase. The ionospheric density on present-day Mars typically peaks below 150 km altitude, well below the exobase \citep{Lee2024}. It should be noted that bulk transport of each ion species, which can be important in the upper thermosphere, is ignored in our thermosphere model.
Using the ion production rate from this model as input, the MHD simulations of ion escape (Section~\ref{sec:ion}) find that the ionosphere density peaks well below the exobase. 
Regardless, the conclusion that \ce{C+} becomes the major ion species in the upper ionosphere is consistent between the thermosphere model and MHD simulations.

\begin{deluxetable*}{ccccccc}
\tablewidth{0pt}
\tablecaption{Total ionization frequencies and Jeans escape rates of major species at the exobase \label{Tab:ionization}} 
\label{Tab:ionizationfrequency}
\tablehead{
  & H & C  & O & O$_2$ & CO & CO$_2$ 
}
\startdata
Ionization frequency (s$^{-1}$) & $7.2 \times 10^{-7}$ & $3.0 \times 10^{-5}$ & $2.4 \times 10^{-5}$ & $1.9 \times 10^{-5}$ & $4.3 \times 10^{-6}$  & $5.9 \times 10^{-6}$ \\
Jeans escape rate (s$^{-1}$) & $2.3 \times 10^{25}$ & $2.8 \times 10^{28}$  & $4.7 \times 10^{28}$ & $1.9 \times 10^{20}$ & $1.0 \times 10^{27}$  & $3.2 \times 10^{19}$  \\
\enddata
\end{deluxetable*}

%% file: 5-hydrodynamic.tex
\section{Hydrodynamic escape}
\label{sec:hydrodynamic}

The PLANETary Ionosphere Thermosphere Tool for Research (PLANET-ITTR) is a comprehensive 1D thermospheric model currently under development at NASA Goddard \citep{Glocer2019}. PLANET-ITTR uses computed or observed extreme and far-ultraviolet (EUV/FUV) stellar fluxes to generate planetary ionospheres from neutral density profiles in a self-consistent manner. It is designed in a modular fashion which allows key physical processes such as hydrodynamics, photochemistry and collisional chemistry, heat conduction, eddy and molecular diffusion, as well as non-local thermodynamic equilibrium heating and radiative cooling, to be activated or deactivated for any specific simulation. The hydrodynamical equations are explicitly included in PLANET-ITTR and this module has been verified against the same Parker wind problem outlined in \cite{Tian2005}; the other modules of the model are currently undergoing completion and validation.

We run PLANET-ITTR for a simplified version of the atmosphere that was generated by the thermospheric model in Section 4. The goal with this PLANET-ITTR run is to assess whether hydrodynamic escape may be occurring. We simulate a single species isothermal atmosphere of neutral oxygen at 3,000 K. This is a representative temperature from the thermospheric model, which reaches 3,000 K by 550 km altitude and is over 4,000 K at 1,750 km.  The surface density for oxygen ($10^{10}~\mathrm{cm}^{-3}$) is also selected based on the results of the thermospheric model. We include only the hydrodynamic equations in this assessment run; there are no other physical processes (e.g., chemical reactions, collisions, heating).

\begin{figure}[h]
\centering
\includegraphics[width=1\linewidth]{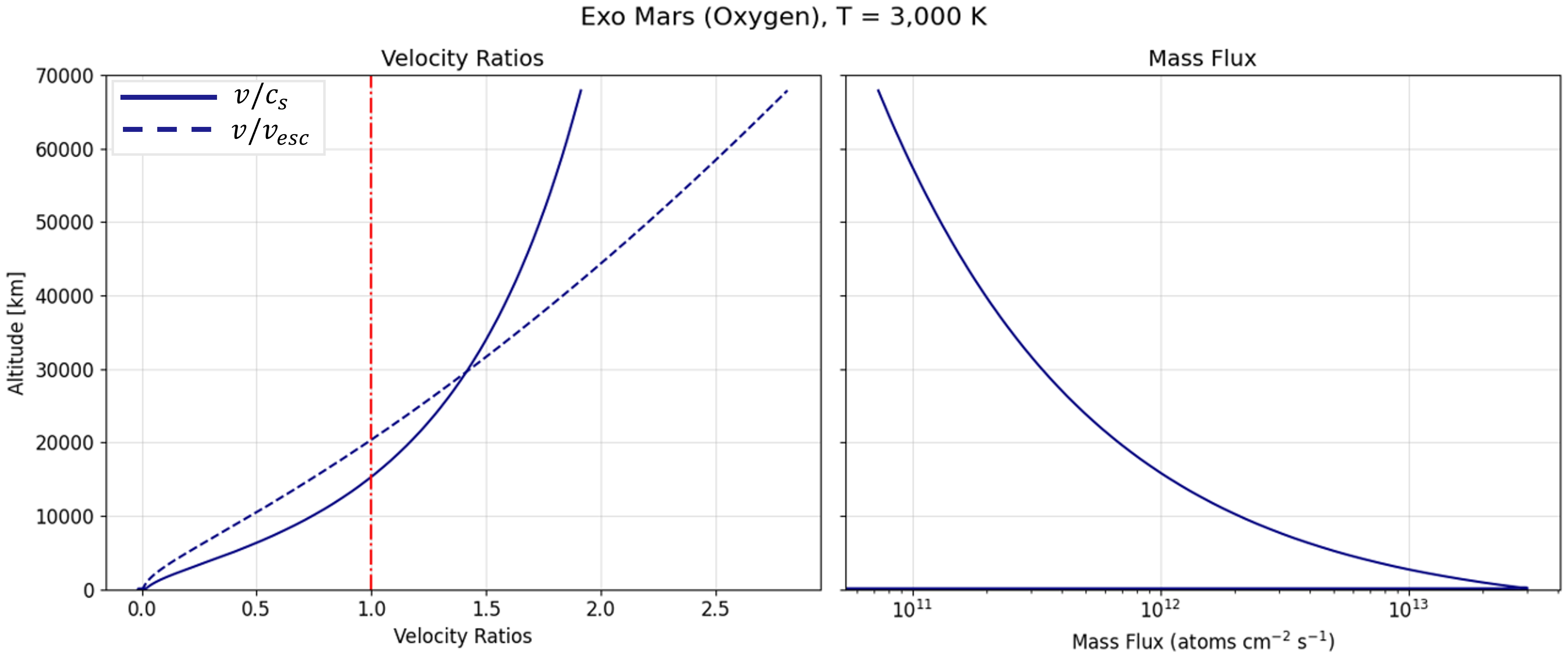}
\caption{ \textbf{Left:} Shown are two velocity ratios as a function of altitude (y-axis). The solid blue line shows the speed of the upward traveling oxygen divided by the speed of sound. The dashed blue line shows the speed of the upward traveling oxygen divided by the escape velocity at each altitude. The atmosphere reaches the sonic point about 4,000 km before it reaches escape velocity. \textbf{Right:} Shown is the mass flux of oxygen atoms at each altitude. The profile is such that, when integrated over a sphere, it is constant at each altitude ensuring continuity. \label{fig:hdyrdyn_esc}}
\end{figure}

If the upward velocity of the neutrals exceeds the sound speed of the gas, the atmosphere can be considered to be hydrodynamically escaping \citep{Tian2005}. Figure~\ref{fig:hdyrdyn_esc} shows that this is the case in the assessment run. On the left-hand plot the $x$ axis shows the velocity ratio (gas velocity compared to escape and sound speeds) and on the $y$ axis is altitude in $km$. At an altitude of about 16,000 km, the velocity of the gas becomes supersonic. The upward-traveling neutral oxygen obtains escape velocity at 20,000 km. We estimate the global oxygen escape rate from the top of the atmosphere to be $4.6\times 10^{31}$ atoms/s using the mass flux (Figure~\ref{fig:hdyrdyn_esc}b) at the upper boundary of the model multiplied by the global surface area ($Q=nvA$, where $n$ is the number density of oxygen, $v$ is the velocity of oxygen, and $A$ is the surface area). 

We compare the estimated escape rate of this model run with the analytical energy-limited escape rate $\dot{M}_{\mathrm{EL}} \approx \eta \pi R_p^3 F_{\mathrm{XUV}}/(G M_p)$ using $\eta=0.1$ and $F_{\mathrm{XUV}}=5~\mathrm{W\,m^{-2}}$ (Eq. 5 in \cite{luger2015}). We find $\dot{M}_{\mathrm{EL}} \approx 1.4\times 10^6\,\mathrm{kg\,s^{-1}}$, which corresponds to an oxygen atom loss rate of $\dot{N}_{\mathrm{EL}} \approx 5\times 10^{31}$ atoms/s; this is in agreement with the model estimate. We further compute the Jeans parameter ($\lambda = G M_p m / (k T r_{exo})$) with $T = 3000~\mathrm{K}$ and an exobase altitude of $2,100 \ \mathrm{km}$ to be $\lambda \sim 5$. A Jeans parameter of order 5 places the atmosphere in a transitional regime where a bulk, hydrodynamic outflow and Jeans-type escaping tail can operate simultaneously. We compute the Knudsen number at the sonic point, using a collisional cross section of $1\times 10^{-19} \ \mathrm{m}^2$, to be $Kn = 0.09$. This also suggests that the atmosphere is still sufficiently collisional at the sonic point to enable hydrodynamic escape flows.

The estimated hydrodynamic escape rate ($\sim 5\times 10^{31}$) is quite high compared to thermally escaping neutral hydrogen ($10^{26}$ to $10^{27}~\mathrm{s}^{-1}$) on present-day Mars \citep{gronoff2020}, and also exceeds present-day heavy-ion escape ($\sim 4 \times 10^{24}~\mathrm{s}^{-1}$) \citep{brain2015, Ramstad2021}. If, indeed, heavy neutral atoms were hydrodynamically escaping at this estimated rate, it would result in a rapid loss of the atmosphere. However, the purpose of this assessment run is only to indicate whether or not hydrodynamic escape might be operating. The assumptions made limit its use for providing quantitative rates with high confidence. This result suggests that the atmosphere is likely to be hydrodynamically escaping, but follow-on studies are required to further constrain the question. Future work involves turning on the remainder of modules in PLANET-ITTR and relaxing the isothermal assumption, which may result in a cooling of the upper atmosphere that could decrease or stop hydrodynamic escape altogether.

%% file: 6-ion.tex
\section{Modeled ion escape} \label{sec:ion}

\subsection{Model description}
Three global multi-species magnetohydrodynamics (MHD) models are used to independently estimate the ion escape from exo-Mars driven by the stellar wind. The three models are BATS-R-US, REPPU-Planets, and MAESTRO. This subsection provides a brief overview of each model.

\subsubsection{BATS-R-US}
The BATS-R-US model, also referred to as the Block-Adaptive-Tree-Solar wind-Roe-Upwind-Scheme, is a versatile and high-performance magnetohydrodynamic (MHD) code that utilizes adaptive mesh refinement (AMR). It solves 3D MHD equations in finite-volume form, employing numerical methods connected to Roe's Approximate Riemann Solver \citep[]{Powell1999,Tóth2012}.
The model has been widely used to study the Mars-solar wind interaction \citep[]{Ma2002JGR, Ma2007GRL, Ma2014JGR} and has been compared with spacecraft observations, including observations from Viking \citep[]{Ma2004JGR}, Mars Global Surveyor (MGS) \citep[]{Ma2014GRL}, and the Mars Atmosphere and Volatile EvolutioN mission (MAVEN) \citep[]{Ma2015GRL, Ma2017JGR, Ma2018GRL, Ma2019JGR}. For this study, the model solves the continuity equations for five ion species in the ionosphere of terrestrial-type planets, including $\mathrm{H^+}$, $\mathrm{C^+}$, $\mathrm{O^+}$, $\mathrm{O_2^+}$, and $\mathrm{CO_2^+}$. Local time stepping is used to enable fast convergence toward a steady state. In the calculation a spherical grid is used, covering a simulation domain from 100 km altitude to 80 $\mathrm{R_M}$ (3400 km), with radial resolution ranging from 5 km near the inner boundary to ~1 $\mathrm{R_M}$ near the outer boundary.  

\subsubsection{REPPU-Planets}
The REPPU-Planets (REProduce Plasma Universe) code is a three-dimensional multispecies single-fluid MHD model, which was introduced in early papers \citep[]{Terada2009Astrobio, Terada2009JGR,Sakai2018GRL, Sakai2021JGR, Sakai2023JGR, Sakai2026MNRAS, Sakata2020JGR, Sakata2022JGR, Nishioka2023JGR}. The model was originally constructed to be applied to unmagnetized objects \citep[]{Tanaka1993JGR} and afterward was adapted for the Earth's magnetosphere and planetary ionosphere \citep[]{Tanaka1998EPS, Terada2009Astrobio, Terada2009JGR}. It implements the finite volume total variation diminishing (TVD) scheme as well as the monotonic upstream scheme for conservation laws (MUSCL) and the van Leer's limiter \citep[e.g.][]{Tanaka1994JCP,Murawski2002}, which obtains the three-order accurate scheme. Numerical fluxes are computed using a linearized Riemann solver. This model can solve the continuity equation for 14 ion species present in the ionosphere of terrestrial-type planets: $\mathrm{CO_2^+}$, $\mathrm{O_2^+}$, $\mathrm{NO^+}$, $\mathrm{CO^+}$, $\mathrm{N_2^+}$, $\mathrm{O^+}$, $\mathrm{N^+}$, $\mathrm{C^+}$, $\mathrm{He^+}$, $\mathrm{H_2^+}$, $\mathrm{H^+}$, $\mathrm{Ar^+}$, $\mathrm{Ne^+}$, and $\mathrm{Na^+}$. A triangle unstructured grid is applied to the model \citep[]{Moriguchi2008JGR}, covering a simulation domain from 100 km altitude to 72 $\mathrm{R_M}$ upstream and 80 $\mathrm{R_M}$ downstream. The spatial resolution of the grid is approximately $3^{\circ}$ in the latitudinal direction and $4^{\circ}$ in the longitudinal direction, with a radial grid spacing of about 4 km near the lower boundary that increases exponentially with altitude. The time interval of the simulation is determined by the Courant-Friedrich-Lewy (CFL) conditions. A detailed description of the model is found in Section 2 of \citet{Sakai2021JGR}.

\subsubsection{MAESTRO}
The MAESTRO (Multifluid Atmospheric Escape Simulations Toward Real elucidatiOn) model is a three-dimensional MHD model developed for simulating solar wind-Mars interactions, described in \cite{Sakata2024}. The model solves the multifluid or multispecies MHD equations in finite difference form, using the semidiscrete second-order central scheme with a minmod limiter \citep{Kurganov2000}. The time step is determined based on the CFL condition. It utilizes the cubed sphere grid \citep{Ronchi1996} that consists of six faces with a quasi-uniform grid coordinate. The angular resolution of the cubed sphere grid is approximately $3^{\circ}$ in this study. The simulation domain is from 100 km altitude to 80 $\mathrm{R_M}$, with radial resolution ranging from about 4 km near the inner boundary to 1.7 $\mathrm{R_M}$ near the outer boundary. A detailed description of the model is found in Section 2 of \cite{Sakata2024}.
For this study, the model performs a multispecies MHD simulation considering five ion species ($\mathrm{H^+}$, $\mathrm{C^+}$, $\mathrm{O^+}$, $\mathrm{O_2^+}$, and $\mathrm{CO_2^+}$).

\subsection{Model Results}
The stellar wind parameters used by all three models are based on orbit averaged results from the AWSOM model as discussed in Section 3.3, with the stellar wind density $\mathrm{N_{SW}} = 67.5 \ \mathrm{cm^{-3}}$, velocity $\mathrm{U_{SW}} = (-602.1, 0.0, 0.01)$ km/s, interplanetary magnetic field $\mathrm{B_{IMF}} = (-15.4,-4.9, 0.04)$ nT, and plasma temperature $\mathrm{T_{SW}} = 4.8 \times 10^5 $ K. For the case presented, the intrinsic planetary magnetic field is set to 0 nT with no crustal magnetic field included. 

\begin{figure}
    \centering
    \includegraphics[width=1\linewidth]{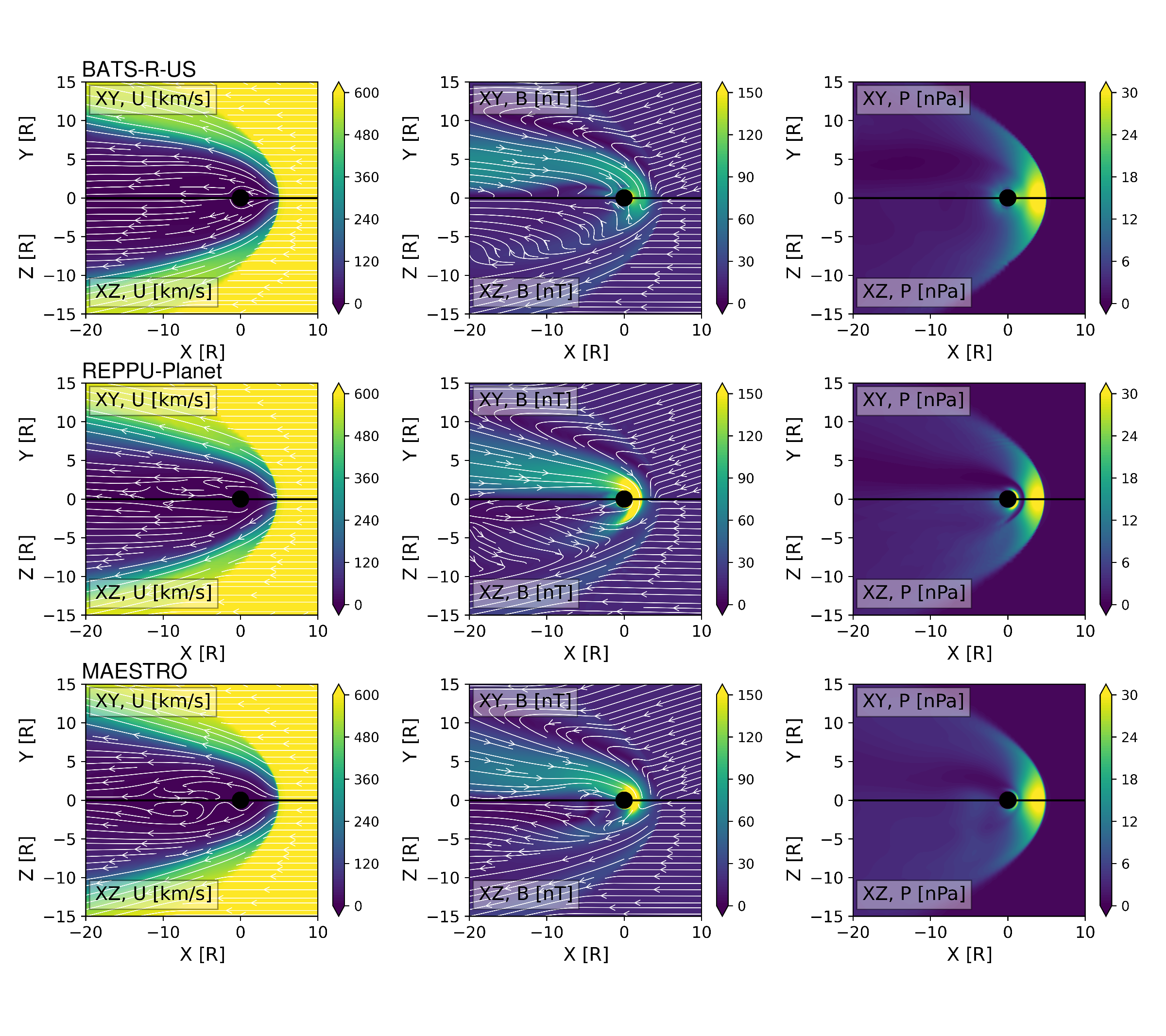}
    \caption{Contour plots for exo-Mars of plasma flow speed (U), magnetic field (B), and plasma thermal pressure (P) in the XY plane (top half) and XZ plane (bottom half) of each panel, from the BATS-R-US (top row), REPPU-Planet (middle row), and MAESTRO (bottom row) models. The white lines with arrows represent streamlines in the U plot and magnetic field lines in the B plot, respectively.}
    \label{fig:MHD_UBP}
\end{figure}

\begin{figure}
    \centering
    \includegraphics[width=1\linewidth]{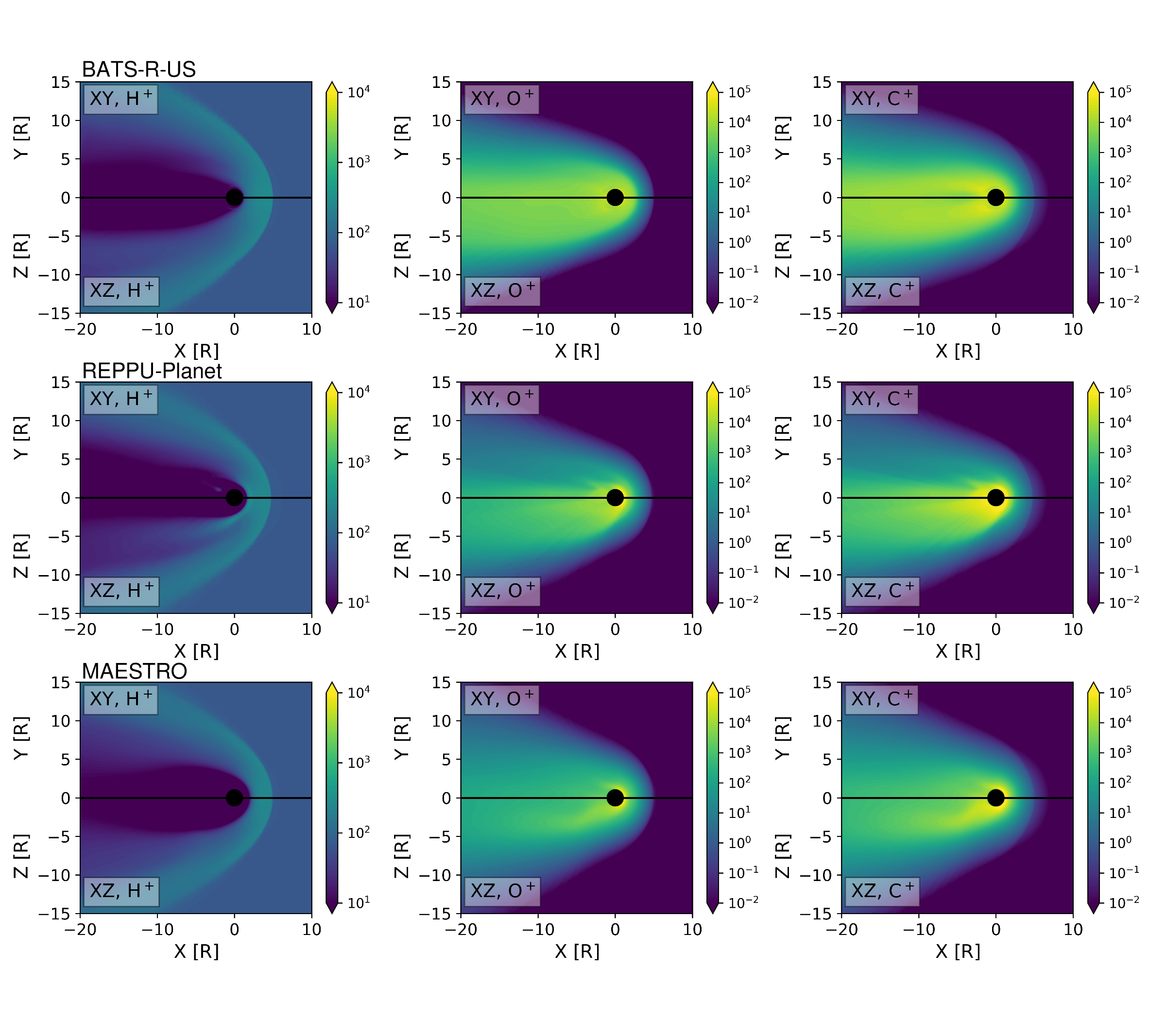}
    \caption{Density distribution for exo-Mars of $\mathrm{H^+}$, $\mathrm{O^+}$, and $\mathrm{C^+}$ in the XY plane (top half) and XZ plane (bottom half) of each panel, from the BATS-R-US (top row), REPPU-Planet (middle row), and MAESTRO (bottom row) models.}
    \label{fig:MHD_densities}
\end{figure}

\subsubsection{Stellar wind interaction with Mars}
The general plasma interaction for a Mars-like planet with stellar wind from Barnard's star is shown in Figure~\ref{fig:MHD_UBP}, which presents contour plots of plasma flow speed (U), magnetic field (B), and plasma thermal pressure (P) in the XY (orbital) and XZ (noon-midnight) planes from the BATS-R-US, REPPU-Planets, and MAESTRO models. The interaction patterns across the three models are generally consistent, with some noticeable differences.
Compared to present-day Mars, all three models predict that the bow shock is much farther away from the planet, approximately 5 $\mathrm {R_M}$ away in the subsolar region, compared to around 1.5 $\mathrm {R_M}$ under current Martian conditions (at solar wind pressure $\sim50$ times smaller than for exo-Mars). This larger interaction region is mainly due to the significantly extended atmosphere for the exo-Mars scenario. Additionally, we note that the simulation results from the BATS-R-US model are based on a steady-state simulation, while both REPPU-Planets and MAESTRO were run in time-accurate mode. A recent model comparison study by \cite{Sun2024ESS}, showed that the BATS-R-US and MAESTRO results agree well under present-day Martian conditions. Comparison of the simulation results of the three different models also highlights that different numerical schemes and detailed treatment could impact the flow pattern in plasma wake, the strength of the magnetic field in the pile-up region, and plasma pressure distribution in the ionosphere, even with the same neutral atmosphere. 

Figure~\ref{fig:MHD_densities} shows the density distributions of three ions around the planet. Among these species, $\mathrm{O^+}$ and $\mathrm{C^+}$ are the most abundant, extending beyond 20 $\mathrm{R_M}$ in the tail region. In contrast, the dominant ions in the current Mars ionosphere are $\mathrm{O_2^+}$ and $\mathrm{O^+}$. 

\subsubsection{Ion escape rates}
The MHD models predict that ion escape rates for exo-Mars are several orders of magnitude higher than those observed for present-day Mars. The ion escape rates from these MHD simulations are summarized in Table~\ref{tab:escape_rate}.
The ion escape rates are integrated at r = 40 $\mathrm{R_M}$ for the BATS-R-US, REPPU-Planets, and MAESTRO models. All three models consistently predict $\mathrm{O^+}$ and $\mathrm{C^+}$ as the dominant escaping species, with loss rates on the order of $10^{28} \ \mathrm{s^{-1}}$. Among them, BATS-R-US tends to predict slightly higher loss rates compared with MAESTRO and REPPU-Planets. Despite modest quantitative differences, the overall agreement indicates robust predictions across independent modeling frameworks.

\begin{deluxetable*}{cccc}
\tablewidth{0pt}
\tablecaption{$\mathrm{O^+}$ and $\mathrm{C^+}$ escape rates at 40 $\mathrm{R_M}$ for each model\label{tab:description}}
\tablehead{
\colhead{Ion Species} & \colhead{BATS-R-US} & \colhead{REPPU-Planets} & \colhead{MAESTRO}
}
\startdata
$\mathrm{O^+}$ & $1.5 \times 10^{28}$ & $4.6 \times 10^{27} $ & $4.6 \times 10^{27}$ \\
$\mathrm{C^+}$ & $1.3 \times 10^{28}$ & $9.4 \times 10^{27}$ & $1.1 \times 10^{28}$ \\
$\mathrm{Total}$ & $2.8 \times 10^{28}$ & $1.4 \times 10^{28}$ & $1.6 \times 10^{28}$ \\
\enddata
\tablecomments{All units are in $\mathrm{s^{-1}}$. The total escape rate for present-day Mars is about $5 \times 10^{24}$ (\cite{Dong2023Icarus}).}

\label{tab:escape_rate}
\end{deluxetable*}

%% file: 7-photochemical.tex
\section{Modeled photochemical escape}
\label{sec:photochemical}

\subsection{Review of Photochemically Escaping O at Present-Day Mars}

A major loss process for oxygen at present-day Mars is photochemical escape.  For this escape mechanism, dissociative recombination (DR) of ionospheric \ce{O2+} ions with electrons produces oxygen atoms, about half of which have speeds greater than the escape speed.  The major ion species at Mars is \ce{O2+} which results from photoionization of the major neutral species carbon dioxide plus some ion chemistry.  

\hspace{2cm} \ce{O2+ + e -> O(3P) + O(3P) + 6.96 eV \hspace{1cm} (0.22)}

\hspace{3.45cm} \ce{-> O(3P) + O(1D) + 5.00 eV \hspace{1cm} (0.42)}

\hspace{3.45cm} \ce{-> O(1D) + O(1D) + 3.02 eV \hspace{1cm} (0.31)}

\hspace{3.45cm} \ce{-> O(1D) + O(1S) + 0.80 eV \hspace{1cm} (0.05)}\\

Branching ratios (from \citet{fox2009}) are shown in the parentheses.  The O atoms for the top two branches have speeds exceeding the escape speed of Mars.

\cite{cravens2017} estimated the escape rate by assuming that each ionizing solar photon absorbed above the exobase would lead to a dissociative recombination reaction, and hence fast O atoms, of which about 25\% could escape.  The 25\% is from roughly the 50\% of the escape channels of the DR reaction that have escape speed and the 50\% of the atoms that are upgoing. The expression obtained for the O escape flux on the dayside was:
\begin{equation}
    F_{esc,O} \cong \frac{f\thinspace I_{CO2}}{4\thinspace \sigma_{O,CO2}}
\end{equation} 

$I_{CO2}$ ($\approx 10^{-6} s^{-1}$ for average solar activity) is the ionization frequency for \ce{CO2} at Mars, $f \approx 0.5$ is an efficiency factor, and $\sigma_{O,CO2}$ ($\approx 1.3 \times 10^{-15} cm^{2}$) is the effective backscatter cross section for O collisions with \ce{CO2}.  

The escape flux from this expression for Mars is $F_{esc,O} \approx 8 \times 10^7 cm^{-2} s^{-1}$.   The global escape rate of O for photochemical escape is then $Q \approx 4 \times 10^{25} s^{-1}$.

\cite{lillis2017} take a different approach and use MAVEN neutral density, ion density, and electron temperature data to determine the DR rate empirically. The results (and comparisons with other methods) depend directly on the O scattering cross section. \cite{lillis2017} obtained an escape rate comparable to that of \cite{cravens2017}.

\subsection{Photochemical Escape of Oxygen for Mars as an Exoplanet}

Overall, the ionization frequency for Mars orbiting Barnard's star is about 30 times greater than for present-day Mars, due to a combination of increased stellar EUV flux and closer orbital distance. Given this, the escape rate for Mars orbiting Barnard's Star with the same neutral composition as present day Mars would be, according to \cite{cravens2017}, about 30 times greater, or $Q \approx 10^{27} s^{-1}$.

However, the neutral composition at our exoplanet Mars is not the same as at present day Mars (see Fig.~\ref{Fig:thermosphere}). The neutral exobase on the exoplanet occurs at an altitude of roughly 2100 km according to our models, with a neutral density of about $10^6$~cm$^{-3}$ (see Section~\ref{sec:thermosphere}). 
The composition at this altitude is dominated mostly by atomic species (O, N, C, CO). The modeled ionosphere looks quite different from that of present day Mars. The ionosphere is dominated by \ce{C+}, unlike present-day Mars. \ce{O+} and \ce{N+}, the two next most abundant species near the exobase, are $\sim$~100 times less abundant. 

Adopting the \cite{cravens2017} photon approximation but scaling the escape rate by the \ce{O2+}/$n_e$ ratio ($\approx$ 0.25) gives an escape rate about 25\% of the previous estimate: $Q \approx 2-3 \times 10^{26} s^{-1}$.

A very crude integration of the dissociative recombination rate (i.e., the \cite{lillis2017} method) and using the atmosphere/ionosphere from Section 4 gives about 2-3 times less: $Q \approx 10^{26} s^{-1}$.

%% file: 8-sputtering.tex
\section{Modeled sputtering escape}
\label{sec:sputtering}

Sputtering process results from direct impact of incident particles (typically ions) with upper atmospheric particles. The collisions transfer kinetic energy to atmospheric particles, which escape. Planetary ions are more likely to cause significant sputtering escape than solar wind ions, for two reasons: (1) planetary ions are more massive, so typically carry an order of magnitude more energy than solar wind protons; and (2) most solar wind protons are deflected in the planet's magnetosphere, and do not directly impact the atmosphere. At present day Mars, sputtering escape rates are typically much lower ($10^{22}$ - $10^{24}$ s$^{-1}$) than escape rates via other processes \citep{leblanc2018, curry2025}. However, sputtering can remove massive atmospheric species, including inert gases such as Argon that are not effectively stripped via any other loss process.

To estimate loss rates via sputtering at Mars orbiting Barnard's star, we first need to estimate the flux of planetary ions impacting the atmosphere, and in particular the energy distribution of the impacting particles. We therefore develop a test-particle model to reconstruct the trajectory of newly ions created from the extended exosphere up to the exobase at 2100 km (Section~\ref{sec:thermosphere}). We focused on the heaviest particles, namely, \ce{O+}, \ce{C+}, \ce{CO2+} and \ce{O2+}. The magnetic and electric fields calculated by BATS-R-US (section 6.1.1) are used. The main difficulty to tackle is the duration of such a calculation since, contrary to Mars’ present conditions, test-particles need to be followed from very large distance from the exobase (up to 3.5 planetary radii) because of the large height scale of the heavy exospheric species (Figure~\ref{Fig:thermosphere}). 

\begin{figure}
    \centering
    \includegraphics[width=1\linewidth]{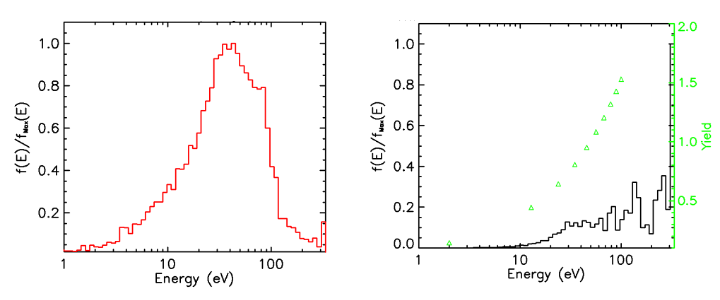}
    \caption{Left: Energy distribution of the O+ ion crossing the exobase. Right: Ejected flux induced by each bin in energy of the precipitating flux in black and corresponding yield value in green.}
    \label{fig:sputtering}
\end{figure}

Figure~\ref{fig:sputtering}, left panel, displays the reconstructed energy distribution of the \ce{O+} ion crossing the exobase. As shown in this figure, the peak in energy of the precipitating ion is around a few tens of eV, the flux decreasing significantly above with a maximum energy of the precipitating particle of few hundred eVs. We used \citet{johnson2000} to estimate the efficiency of the precipitating ion to erode the atmosphere, which is expressed as a yield equal to the ratio between the number of ejected particles with escape energy divided by the number of precipitating particles. \citet{johnson2000} calculated such yield for O atoms in function of the incident energy. In the right panel of Figure~\ref{fig:sputtering}, we displayed the yield values, corrected to take into account the altitude of the exobase (200 km in \citet{johnson2000}). As shown in Figure~\ref{fig:sputtering}, for the range of energy of the precipitating particles (below 100 eV), the typical yield value is lower than 1, meaning that Mars accretes more particles than it loses. Similar calculations and results have been performed for \ce{CO2+}, \ce{O2+} and \ce{C+}. For the present Mars, the typical flux of precipitating particles also peaks at a few tens of eV but is much more extended to higher energy (up to 10 keV).

These results suggest that an extended exosphere as simulated in Section~\ref{sec:thermosphere} leads to a very extended magnetosphere and does not allow pick-up planetary ions to be sufficiently accelerated before reaching the exobase. In another way, an extended exosphere protects an atmosphere from sputtering. Such conclusion is actually consistent with \citet{martinez2020}, who used MAVEN observations that high solar dynamic pressure enhances the flux of ion precipitating into Mars’ atmosphere.\\

%% file: 9-synthesis.tex
\section{Synthesis of escape rates and comparison to present-day Mars}

Table~\ref{Tab:synthesis} summarizes the atmospheric escape rates for exo-Mars computed or estimated in this work via the different escape processes, along with estimates of escape rates via each process from present-day Mars. The rates of Jeans and hydrodynamic escape, ion escape, and photochemical escape are all larger for exo-Mars than for present-day Mars. The sputtering escape rate is likely to be smaller, but sputtering rates are dwarfed by the other processes. Therefore, we conclude from the table that the escape rate for exo-Mars would be significantly greater than for present-day Mars. 

\begin{deluxetable*}{lll}
\label{Tab:synthesis}
\tablewidth{0pt}
\tablecaption{Modeled/estimated escape rates for Mars orbiting the Sun and Barnard's star}
\tablehead{ \colhead{Process} & \colhead{present-day Mars} & \colhead{exo-Mars} }
\startdata
Thermal escape & $10^{26}$ - $10^{27}$ (H) & $10^{19}$ - $10^{28}$ (O, C, CO, H, \ce{02}, \ce{CO2}) \\
Hydrodynamic escape & $\sim$0 & $5 \times 10^{31}$ (O)\\
Ion escape & $4 \times 10^{24}$ ($\mathrm{O^+}$,$\mathrm{O_2^+}$) & $1-3 \times 10^{28}$ ($\mathrm{O^+}$,$\mathrm{C^+}$) \\
Photochemical escape & $4 \times 10^{25}$ (O) & $1-3 \times 10^{26}$ (O) \\
Sputtering & $10^{22}$ - $10^{24}$ (\ce{O}, \ce{CO2}, \ce{Ar}, \ce{C}, \ce{N}, \ce{N2}, …) & $\sim$0 \\
\enddata
\tablecomments{All units are $\mathrm{s^{-1}}$. 
}
\end{deluxetable*}

Taking a conservative estimated escape rate for oxygen of $\sim6 \times 10^{28} \ \mathrm{s^{-1}}$ (combining the modeled rates from Jeans escape, ion escape, and photochemical escape, but omitting hydrodynamic escape), we can compute the time necessary to deplete all of the oxygen incorporated into atmospheric $\mathrm{CO_2}$. A $\sim$ 7 mbar pure $\mathrm{CO_2}$ atmosphere similar to present-day Mars contains roughly $7 \times 10^{41}$ oxygen atoms. At the escape rate given above, it would take about 350,000 years to remove the entire atmosphere. If the atmosphere instead had a surface pressure of 1 bar (a median value postulated for early Mars by \citet{jakosky2018}) but was otherwise the same in every respect, it would take approximately 50 million years to remove the atmosphere. We have considered quiescent stellar conditions in this work; incorporation of the effects of stellar flares and Coronal Mass Ejections (CMEs) would likely increase time-integrated escape rates further, reducing the atmospheric retention timescale even more \citep[e.g.][]{france2020, amaral2025}. If hydrodynamic escape is indeed occurring, as indicated in Section~\ref{sec:hydrodynamic}, then the retention timescale would be reduced yet again, and significantly. We conclude that it is unlikely that observations of any given Mars-like planet orbiting a star similar to Barnard's star near the Habitable Zone would occur during a period that it possessed a secondary atmosphere. 

Geological and geophysical processes should supply volatiles from the exo-Mars interior. Such processes have not been considered explicitly in the modeling exercise presented here. Estimates of outgassing of \ce{CO2} from the interior of present-day Mars are on the order of 1 bar \citep[e.g.][]{grott2011}. \citet{jakosky2023} argue that outgassing occurred primarily in the first few hundred million years, so that our estimate above of 50 million years to remove 1 bar of atmosphere conveniently includes the completion of outgassing as a starting point for the estimate for any M dwarf star older than a few hundred million years. Outgassing may outpace escape during the first few hundred million years, though young M dwarf stars should be significantly more active than Barnard's star, which is roughly 10 billion years old. The escape rate in this time period is therefore likely to be larger than estimated in this work. 

As pointed out by \citet{jakosky2025}, a planet may sequester atmospheric gas in surface or subsurface reservoirs, and release it later in the planet's history. In the case that 1 bar of atmosphere is sequestered and later released, then it may be possible to observe an atmosphere for 50 million years late in the planet's history. For the case of Barnard's star, this period represents 0.5\% of the star's history, corresponding to at most a 0.5\% chance of observing an atmosphere at any given time. For younger stars the likelihood increases. It may be possible that the later release of atmosphere occurs at low rates sufficient to balance the escape rate, allowing an atmosphere to be retained. Such periods should also represent a small fraction of the star's history. For example, if 7 mbar of atmosphere were released every 350,000 years (the time required to remove 7 mbar by our calculations), the 7 mbar present day surface pressure of Mars could be maintained. It would take 50 million years to release 1 bar of atmosphere at this rate. However, detailed modeling of atmospheric source processes is required to be sure of these estimates, or to consider other scenarios (such as slow release of more than 1 bar of atmosphere).

Recently, the detection of four sub-Earth planets have been reported orbiting Barnard's star \citep{basant2025}. All four planets orbit well interior to the exo-Mars scenario considered in this work, by a factor of 2.3-4.6. Assuming that the escape is energy-limited, this should lead to a considerable increase in escape since the EUV energy received by the planet scales as the inverse square of the orbital distance (i.e. by a factor of 5.3-21). However, the planetary masses are larger than for exo-Mars by a factor of 1.6-3.6. The larger gravity should inhibit escape; the escape velocity for atmospheric particles is greater by the same factor, and the energy required for escape is greater by a factor of 3-13. It is difficult to quantitatively scale our results for exo-Mars to different orbital distances and planetary masses without re-running all of our models. Instead, we argue that it is  unlikely that any of these planets possess a secondary atmosphere. This is because the energy required for escape increase by a smaller (or perhaps comparable) amount than the energy from EUV photons that drives escape. Exo-Mars loses atmosphere very rapidly, and it is difficult to imagine that the four planets would lose atmosphere significantly more slowly than exo-Mars. Primary atmospheres seem similarly unlikely, since primary atmospheres are comprised of hydrogen and helium, which are lighter than $\mathrm{CO_2}$ and thus should escape more easily, and were likely removed much earlier in the star's evolution when the stellar XUV flux and wind rates were $\sim$ 100 times larger \citep[e.g.][]{pineda2021, pass2025}.

%% file: 10-discussion.tex
\section{Discussion}

We have presented estimates of atmospheric escape rates via five different escape processes for a Mars-like planet orbiting Barnard's star, a relatively inactive M dwarf star. We find that escape rates for all processes except sputtering increase relative to present-day Mars if we place the planet at an orbital distance where it would receive the same total stellar photon flux. The magnitude of the increase and the inactivity of Barnard's star relative to other M dwarf stars leads us to infer that secondary atmospheres on Mars-sized planets orbiting M dwarfs at distances near the Habitable Zone are unlikely to be observed. This is consistent with the results of attempts to observe atmospheres of M dwarf planets to date \citep{kreidberg2025}. 

The escape rates presented in this work should be taken as rough estimates, though we are confident in the overall trends relative to the escape rates for present-day Mars. There are many challenges to accurately modeling atmospheric escape from exoplanets that it is important to be aware of. We list some of the challenges that surfaced during the course of our work:
\begin{itemize}
    \item Stellar EUV inputs for Barnard's star are not directly measured. Instead they are computed based on measurements at shorter and longer wavelengths than EUV. This Differential Emission Measure technique has shown to be effective at providing reasonable EUV fluxes for a star, but the fluxes can be uncertain by factors of a few \citep[e.g.][]{france2022}.
    
    \item Stellar wind and interplanetary magnetic field inputs in Section~\ref{sec:inputs} are modeled based on the surface magnetic field of the star. Stellar magnetograms are not available for Barnard's star, so a proxy star was used. Additionally, orbit averages were taken of the quantities used to drive magnetospheric models. These quantities vary significantly over the course of an orbit, and certainly will drive variations in escape rates.
    
    \item The thermosphere model in Section~\ref{sec:thermosphere} is one-dimensional. This leads to an assumption of spherical symmetry for the thermosphere when computing Jeans escape rates. A Mars-sized planet orbiting 0.087 AU from its star is likely to be tidally locked, making the assumption of spherical symmetry problematic. Because the model is one-dimensional, an eddy diffusivity profile was required. The value of the eddy diffusivity is poorly constrained even for present-day solar system planets.
    
    \item The ion escape models in Section~\ref{sec:ion} compute escape rates that differ by factors of up to 3, and differ in which escaping species is dominant. These models have different grids, numerical schemes, and implementations of the lower boundary conditions. It is not possible to determine which of the three models provides the `best' estimate of atmospheric escape rates.
    
    \item The photochemical escape estimate in Section~\ref{sec:photochemical} assumes that oxygen escape rates scale linearly with the stellar EUV flux. While this has proven to be a good scaling law for present-day Mars, it may not hold for the more extreme EUV flux from Barnard's star.
        
    \item The sputtering estimate employs an MHD model to reconstruct the trajectory and acceleration of the precipitating particles to the atmosphere. Hybrid models, which capture the kinetic motion of ions, are clearly better adapted to track the acceleration and trajectory of the most energetic pick-up ions.

    \item The escape processes considered in this work were considered independently, using separate models and approaches. This is to be expected given the different physical processes that are operating. But treating the processes independently likely omits coupling between the processes, and may result in "double-counting" of escaping particles (e.g. ions escaping in the magnetospheric models may be counted as neutrals in other models).
 
\end{itemize}

These challenges can be organized into five categories. The first is knowledge of the necessary inputs for modeling atmospheric escape. This is a challenge for many aspects of exoplanet modeling, not limited to atmospheric escape models. To mitigate this challenge it might be prudent to assume a range of input conditions, though this can lead to significant computational expense if models have to be run many times. Instead we chose a combination of intermediate and `best guess' values, and remain mindful that escape rates can vary depending upon the inputs.

A second category is the use of 1D vs 3D models. 1D models are simpler to run, and may be more consistent with the lack of detailed information about the three dimensional structure of exoplanet atmospheres, including day-night and equator-to-pole variations. In the absence of this information, 3D models are free to use different assumptions about boundary conditions, leading to different results. 1D profiles cannot capture the longitude and latitude variation in exoplanet atmospheres - especially for tidally locked planets.

A third category is the use of any individual model to estimate escape rates. Every model used in this work has been validated against observations from planets in our solar system. Yet comparable models (e.g. the magnetosphere models in Section~\ref{sec:ion}) give results that vary by factors of several. Early in our analysis, the differences between comparable models was even greater. Only through discussion and iteration did it become apparent that the different models were employing different assumptions about initial or boundary conditions. This serves as a caution about the interpretation of the results of any single model that is being applied to a situation where it has not been previously validated. The competing models used in this work agreed generally that escape rates were significantly elevated for exo-Mars, but disagreed about the value of the escape rate.

A fourth category is assumptions about scaling laws. Scaling laws can provide a quick and useful means of estimating escape rates without resorting to detailed modeling. But caution must be taken to not apply scaling laws outside of the range of conditions where they are applicable.

A fifth category is balancing the physics captured by models with their computational complexity and convenience. While a hybrid plasma model would better capture the trajectories of incident particles that cause sputtering, MHD models are quicker to run and were already being used to compute ion escape.  

Despite these challenges, we believe that the general results presented here are robust: atmospheric escape rates would be significantly elevated for exo-Mars relative to present-day Mars. Jeans escape rates would increase by multiple orders of magnitude (and include more species than hydrogen). Ion escape would increase by nearly four orders of magnitude (and include significant carbon escape). Photochemical escape of oxygen would increase by factors of 3-8. Uncertainties about input conditions or model assumptions seem unlikely to change the overall inference that escape would greatly increase. Because we have chosen a secondary atmosphere on a planet orbiting an old, inactive M dwarf during quiescent conditions, we further argue that Mars-like planets around any M dwarf star are unlikely to retain atmospheres for significant periods of time.